\documentclass[conference,slashbox]{IEEEtran}

\IEEEoverridecommandlockouts
\usepackage{cite}
\usepackage{amsmath,amssymb,amsfonts}
\usepackage{algorithmic}
\usepackage{graphicx}
\usepackage{textcomp}
\usepackage{xcolor}
\usepackage{slashbox}
\usepackage{soul}
\usepackage{amsmath}

\newcommand{\varB}[1]{{\operatorname{\mathit{#1}}}}

\def\BibTeX{{\rm B\kern-.05em{\sc i\kern-.025em b}\kern-.08em
    T\kern-.1667em\lower.7ex\hbox{E}\kern-.125emX}}
\begin{document}
 
\title{A Fair model of identity information exchange Leveraging Zero-Knowledge\\
 }
\author{
    \IEEEauthorblockN{1\textsuperscript{st} Mohsen Rahnamaei}
\IEEEauthorblockA{\textit{Department of Computer Science and Engineering } \\
\textit{Amirkabir University of Technology}\\
Tehran, Iran \\
Rahnamaei@aut.ac.ir}
 \\
\IEEEauthorblockN{2\textsuperscript{nd} Saeid Tousi Saeidi}
\IEEEauthorblockA{\textit{Department of Computer Science and Engineering } \\
\textit{Amirkabir University of Technology}\\
Tehran, Iran \\
Stoosi@aut.ac.ir}
\and
\IEEEauthorblockN{3\textsuperscript{rd} Siavash Khorsandi}
\IEEEauthorblockA{\textit{Department of Computer Science and Engineering } \\
\textit{Amirkabir University of Technology}\\
Tehran, Iran \\
khorsandi@aut.ac.ir}
 \\
\IEEEauthorblockN{4\textsuperscript{th} Mehdi Shajari}
\IEEEauthorblockA{\textit{Department of Computer Science and Engineering} \\
\textit{Amirkabir University of Technology}\\
Tehran , Iran \\
mshajari@aut.ac.ir}
 }
 \maketitle
 
\begin{abstract}
    Many companies use identity information
    for different goals. There are a lot of
    market places for identity information.
    These markets have some practical issues
    such as privacy, mutual trust and fairing
    exchange. The management of identity
    information is one of the most important
    applications for blockchain, for which
    researchers have proposed a large number
    of models.  In the present paper, an attempt has
     been made to solve the problems that mentioned
      earlier to exchange identity information on the blockchain. 
      By using the game theory we propose a fair model of selling 
      authorized identity information in an environment that include 
       untrusted parties. Moreover we employ ZK-SNARK to protect users’ privacy. 
       Also we use proxy re-encryption to record these information in IPFS.
\end{abstract}
\begin{IEEEkeywords}
    Blockchain, Privacy, ZK-SNARK, Smart contracts, Proxy re-encryption, Game theory, Identity information management
\end{IEEEkeywords}

\section{Introduction}\label{introduction}
Researchers in the field
of blockchain technology has introduced several
applications for blockchain . One of the most
important of these applications is the identity
information management \cite{kuperberg2019blockchain}
. Citizens need to share
their information with others in a certified and
accredited manner. This action allows them to take
advantage of different social services. However,
misuse of this information causes privacy violations \cite{goodwin1991privacy}.

Blockchain is a distributed database on which the accuracy
of the stored data can be guaranteed \cite{swan2015blockchain}. Researchers usually use
this technology for storing metadata. The architecture and
the way of employing this technology, along with some concerns about privacy and fairness ,
is the main challenge for researchers. In the present article, a new model
has been presented to maintain and enhance users’ privacy.

The use of zero-knowledge proof algorithms and other encryption algorithms
is a common tool in enhancing the privacy of blockchain-based applications.
The proxy re-encryption \cite{qin2016survey} algorithms used in the proposed model. A complete description of the function of these algorithms is presented in
Section \ref{basicConcepts:proxyReEncryption}.

In the present study, a model has been implemented by employing these algorithms.
In this model, the selling of identity information is managed by using the peers
as intermediaries in the Ethereum blockchain  \cite{wood2014ethereum}.

Our contribution is a blockchain-based model that helps an entity to sell his/her
identity  information in a fair way and protect his/her privacy.

We  proposed a dynamic game for this exchange  that in the Nash action profile both
owner and buyer properly deal with each other, we can see  a fair and an accurate
trade in this statement. We use smart contract to eliminate central trusted third party
and enforce game rule.In this game the buyer must prove   that he/she get particular data,
however, privacy rights prevent him/her to publish clear data, in order to solve this  problem
we employed ZK-SNARK \cite{groth2016size}.

The article structure is as follows: In Section \ref{relatedWork}, we will explore some related works. In Section \ref{basicConcepts:blockchainAndSmartcontract}, we will review some researches and will discuss their weakness. Section \ref{basicConcepts} consists of basic concepts. We will
review blockchain technology and smart contract. In Section \ref{basicConcepts:privacyAndBlockchain}, we review some concepts about privacy in the blockchain.
We review the basic concepts about zero-knowledge proof in \ref{basicConcepts:ZKP}. The proxy re-encryption algorithm is explained in Section \ref{basicConcepts:proxyReEncryption}.
In the next section, we describe our proposed solution. In \ref{proposedSolution:proposedModel}, we present the basic concepts about our model. Also, in \ref{proposedSolution:valdationTheAccuracyByGameTheory}, we discuss
on the validity of our solution. Finally, in \ref{conclusion}, we describe the conclusion.

\section{RELATED WORK}\label{relatedWork}
In \cite{manzoor2019blockchain} authors develop smart contract that
entities  can employ to trade IOT data. They
use proxy re-encryption to improve scalability
in this market. The confidentiality of data
guaranteed with proxy re-encryption algorithm
and the financial transactions are
automatically managed with smart contract.
In this paper, writers supposed that all
entities are trustable, in other word,
they supposed the owner of information send
his/her data to the buyer without any
dishonesty. Our model develop the possibility of trading
data in a situation where buyer and seller don’t need any trust
on each other. Notably, we have to emphasize that in each exchange
of goods, fairness is very important, in another word seller must get
the money in exchange for goods and similarly for the buyer must get goods
in exchange of money.

There are some researches that trying to improve
fairness in the blockchain based trading \cite{bigi2015validation} \cite{zimbeck2014two} \cite{asgaonkar2019solving}.The basic approaches
to overcome fairness problem is Atomic Swap \cite{asgaonkar2019solving}.Bigi et al. \cite{bigi2015validation} use normal
form game to work out fairness problem. In \cite{bigi2015validation}, authors assume sent goods are
accurate and authentic, while owner may send defective goods. In this game if the
owner send defective good the buyer prefer to be dishonest and verify the transmission,
while he get inaccurate good, because he want to increase his/her payoff.

There exist different studies to facilitate trading data on blockchain \cite{juels2016ring} \cite{manzoor2019blockchain} \cite{zimbeck2014two}.Ari et al. \cite{juels2016ring} Published
one of the first paper that proposed criminal smart contract. These contracts employ to handle financials transactions
for criminals. For example leakage of secrets, key compromise and “calling-card” crimes, such as assassination. Their
propose solution for document leakage is sophisticate in comparison with our solution. We use game theory and ZK-SNARK to
overcome this problem.

\section{BASIC CONCEPTS}\label{basicConcepts}
\subsection{Blockchain and Smart Contract}\label{basicConcepts:blockchainAndSmartcontract}
Blockchain is a transparent distributed database
that everyone can use, and at the same time,
no one has control over it. This database consists
of a set of blocks connected by hash algorithms. The connection
of blocks improves the accuracy of the information contained in
the blocks. These blocks are updated by network peers. None of the
peers has absolute control over the information recorded on the blockchain,
and attempts have been made to use motivational engineering algorithms so that
the peers act in such a way that the blocks to be updated \cite{swan2015blockchain}.Storing metadata on
the blockchain assist in ensuring the accuracy of these information related to this
metadata, which has been stored in other centralized and distributed storage.

Ethereum smart contracts, as a distributed programs on peers available in the Ethereum
blockchain, allow programmers to make sure their applications are running correctly.
With smart contracts, it is possible to build applications and projects to have the
capability of permanently functioning without intermediaries and corruption. Even a smart
contract programmer cannot change the smart contract code registered in the blockchain \cite{chris2017introducing}.
\subsection{Privacy and Blockchain}\label{basicConcepts:privacyAndBlockchain}
The research on blockchains’ privacy is divided into two general categories. In the first category,
attempts are made to alter the blockchain structure using encryption algorithms in a way that privacy
to be improved. These studies often introduce new structures by applying encryption algorithms \cite{dasgupta2019survey}.

The second category is related to research, in which the blockchain is used to improve privacy in other
applications \cite{nabi2017comparative}. In the second category, some of the models proposed for identity information management
on the blockchain, only stores metadata on the blockchain. In these models, the primary information is stored
on centralized or distributed storage, and the output of the hash function is placed on the blockchain. Since it
is challenging to make changes to the information recorded on the blockchain, any unauthorized changes to the information
stored in the distributed database will be recognizable.

\subsection{Zero knowledge proof}\label{basicConcepts:ZKP}
One of the big problems in privacy was proving something
which prover didn’t want to reveal any information about real things,
for example, Alice wants to prove Bob, that she knows some secrets, but
she doesn’t prefer to show that to Bob. Zero-knowledge proof devised to resolve
this problem .This notion  proposed in 1989 by MIT researchers Goldwasser et al. \cite{goldwasser1989knowledge}. They
explained the concept of knowledge, and how can convincingly prove the statement. In all zero-knowledge
protocols such as ZK-SNARK, we can see completeness, soundness and zero-knowledge.
\begin{itemize}
\item Completeness: If the statement is correct, then the verifier will convince by an honest prover.
\item Soundness: If the statement is not true, any prover can’t convince the honest verifier, except some small probability.
\item Zero knowledge: Prover doesn’t disclose the private information to verifier.
\end{itemize}

In this paper, the ZK-SNARK(Zero-Knowledge Succinct Non-Interactive Argument of Knowledge) is leveraged to preserve
privacy in data transmission and proving of that. For implementation phase, Zokrates \cite{eberhardt2018zokrates} used to provide verifier smart
contract and proof. This library provides high level language to generating proofs of computation and verification in Solidity.

ZK-SNARK requires initial trusted setup phase which generates the common reference string and it must be done by a trusted entity.
This is a challenging phase of the protocol. Using a trusted entity is not irrational, but there is another solution which can use to
eliminate relying on the centralized entity. Zcash generated ZK-SNARK parameters by using the trusted setup ceremony(multiparty computation)\cite{zcashsksnark}.

\subsection{Proxy Re-Encryption}\label{basicConcepts:proxyReEncryption}
Proxy re-encryption algorithms \cite{qin2016survey} are a set of algorithms that
allow an unreliable intermediary to re-encrypt an encrypted text by another
key. The unreliable intermediary can perform this process without knowing anything
about the data. Proxy re-encryption algorithms are similar to the public key encryption \cite{boneh2015graduate},
in which a pair of public-private keys are used to encrypt and decrypt the data.

\begin{figure}[htbp]
	\centerline{\includegraphics[width=0.4\textwidth]{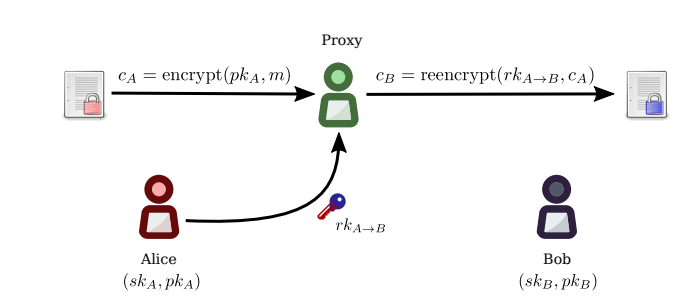}}
	\caption{Proxy re-encryption protocol \cite{nunez2018umbral} }
	\label{fig:proxyfig}
\end{figure}

The proxy re-encryption eliminates the requirement of re-encryption.
The proxy receives users’ encrypted message, then re-encrypts and sends
it to an entity. This feature prevents the repetition data encryption by a
symmetric key. In Fig.~\ref{fig:proxyfig}, the process of proxy re-encryption is described.

In Fig.~\ref{fig:proxyfig}, $sk_{A}$ is referred to the private key of data sender (Alice), $pk_{A}$ is the public
key of data sender,  $sk_{B}$ implies to the private key of data receiver (Bob), $pk_{B}$ is the public key of
data receiver, and finally, $rk_{A}$ is the proxy re-encryption key.
Assume that Alice wants to send a message to Bob. This message is encrypted by $pk_{A}$ and provided to the
proxy. When Bob wants to access this information, he sends his request to the proxy. At this stage,
Alice uses $sk_{A}$ and $pk_{B}$ to generate $rk_{A}$  and provides this key to the proxy. By employing this key, the proxy
converts the previously encrypted message into a new encrypted message and provides it to Bob without
being informed of the original message's content. Bob can also decrypt this message using $sk_{B}$.

This feature makes the untrusted intermediary unable to access information perform the proxy re-encryption at the same time.
The proxy re-encryption algorithm includes five functions: generating asymmetric keys, encryption, generating the
proxy re-encryption key, proxy re-encryption, and decryption. In the following, it is presented that how each
of these functions works \cite{ateniese2006improved}.
\begin{itemize}

\item $Generating \; asymmetric \; keys$: In this
function, the parameters and the pair
of primary keys of the design are generated.
First, a cyclic group should be created,
which is generated as $<g>=\mathbb{G} $ with the order $q$  and
the generation $g$. This group is used to generate
asymmetric keys. In the next step, a random value
called $a$ from $\mathbb{Z}^{*}_{q}(a \in \mathbb{Z}^{*}_{q})$ is selected as the $sk_{A}$ private key, and
then, the public key can be generated as $pk_{A}=g^{a}$  by
applying $a$ and the generator of the group.
Moreover, the value of $Z=e(g,g)$ which is global value, must be created for use in subsequent functions.

\item $Encryption$: After generating the pairs of keys,
the information can be encrypted using a random
value. This function firstly selects a random
value named $r$ from $\mathbb{Z}^{*}_{q}$ . Then the value of $m \in \mathbb{G}_{2}$, which
is the selected message, is encoded using the
random value $r$  and the value $Z$  as $C_{a}=(Z^{r}.m,g^{ra})$

\item $Generating\; the\; proxy\; \varB{re-encryption}  \; key$:
The proxy
re-encryption key dedicated to Bob $(b)$ can be generated
using this function. The public key related to Bob
and the private key of Alice $(sk_{A})$  is used to perform
this action. The public key of Bob $ (pk_{B}) $ is powered by $1/a$ ,
and this value is used as the proxy re-encryption
key from Alice to Bob.

\item $Proxy \; \varB{re-encryption}$: The input values of
this function consist of an encrypted message and
the proxy re-encryption key, and the purpose of
this function is to proxy re-encryption of messages
without disclosing the information. For this purpose,
the following process is applied to the encrypted
message.

$C_b=(Z^{r}.m,e(g^{ra},rk_{a}))=(Z^{r}.m,e(g^{ra},g^{b/a}))=(Z^{r}.m,Z^{rb})$

\item $Decryption$: After running re-encryption function on the
message by a re-encryption key, Bob can use this function
and his private key to decrypt the encrypted message.

\end{itemize}

\section{PROPOSED SOLUTION}\label{proposedSolution}
The main problems with previous researches
are the lack of the privacy and fairness in
the data exchange, such as the proposed model
on the identity exchange via blockchain. Since
entities are untrusted in the real world, we propose our solution
based on the game theory and ZK-SNARK. To protect the confidentiality
of data and reduce the computation, we use proxy re-encryption algorithm.
In our solution, the sender encrypts the information only once using
the re-encryption proxy algorithm.

In the following, we will describe how this protocol works. It is noteworthy
that this model can be used in all markets where buy and sale of certified data
are made.

\subsection{Proposed model}\label{proposedSolution:proposedModel}
In the present paper, a new model based on a smart contract is introduce for selling stored certified data on decentralized databases such as
IPFS \cite{benet2014ipfs} and SWARM \cite{bashir2017mastering}.
In order to further explain this protocol, we present the following
scenario. Assume that the seller wants her information involving all kinds of certified
data related to her identity to be placed on a decentralized database such as IPFS (Step 1).
The seller must encrypt her information by her public key and then place the cipher on the distributed
database. Now assume that the buyer wants to buy this information from the seller.

\begin{figure}[htbp]
	\centerline{\includegraphics[width=0.48\textwidth]{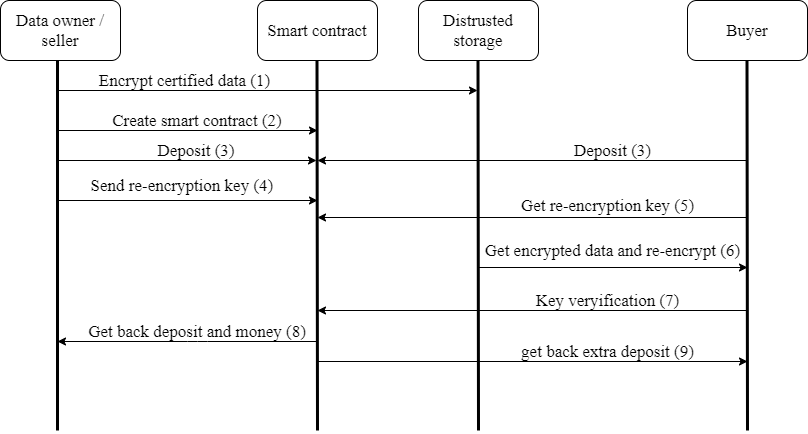}}
	\caption{The function of the proposed protocol.}
	\label{fig:thefunctionvalidation}
\end{figure}

After agreeing on the price of this information,
the seller creates a smart contract on the Ethereum blockchain (Step 2).We assume none of the seller and the buyer entities are trustable in this
scenario. Also, the amount of deposit is calculated in a way that incentives
both parties to perform correctly; the amount of deposit required to prove the validity
has been examined in Section \ref{proposedSolution:valdationTheAccuracyByGameTheory}. The seller must send the encryption key to the smart
contract after sending the deposit.This key calculate with proxy re-encryption algorithm, so no one can use it, except the buyer.To decrypt information, the buyer must use his private key. In fact, the decryption key depends on the re-encryption key and the buyer private key. Steps 3 to 5 in Fig.~\ref{fig:thefunctionvalidation} show this process.

After performing this action, the smart contract transmits the proxy re-encryption key to the buyer (Step 5). In the next step, the buyer must take this information from the decentralized database and decrypt it (Step 6).As we mention earlier, the decryption key calculate with the buyer's private key and re-encryption key. If the buyer
accesses to sender's certified information, he will notify the validity of that to the smart contract with ZK-SNARK(Step 7).

Finally, the seller  withdraws the price of data from the smart contract, also the smart contract get back deposit to the buyer and the seller
. Steps 8 and 9 in  Fig.~\ref{fig:thefunctionvalidation} show
the process of these steps in protocol.

If each entity fails to properly perform its related function
in steps 3, 4, and 5, the deposits will be blocked by the smart contract and will not be refunded to any of the members . This policy prevents buyer and seller to have dishonest actions. The validity  of this model has been
proven by game theory in Section \ref{proposedSolution:valdationTheAccuracyByGameTheory}.

\subsection{Validating the model by game theory}\label{proposedSolution:valdationTheAccuracyByGameTheory}
In this section, we will evaluate the amount of deposit required for the validity of this model. In this game, we have two
players, the buyer, and the seller. The seller has two main actions. First The seller send the valid data; The second action assumed as an action when the seller doesn’t send anything or send invalid data. Also,
the buyer has two main actions.First, the buyer can confirm the data that he gets from the seller.  The second action is
when the buyer does not approve data. The utility function for each player defines how much they get at the end of the exchange. It is essential to denote,
information has value, and these values are counted as $v_{s}$ or $v_{b}$. The utility function comes from the value of the data, price of the data which has been
paid, and the amount of deposit. The parameters in Table.~\ref{tab:strategicform} are as follows and each cell of the table shows as
this format ( buyer payoff,seller payoff ):

$d_{s}$: Sellers’ deposit amount

$d_{b}$: Buyers’ deposit amount

$v_{s}$: Value of the data for the seller

$v_{b}$: Value of the data for the buyer

$c$: Price of information

We assume that $v_{s}$, $v_{b}$ and $c$  is equal but as we mentioned  earlier  the amount of deposit is higher than $c$. In each cell of Table ~\ref{tab:strategicform}, we show the level of benefit of players in different condition.
In this table, the Nash equilibrium \cite{myerson2013game} modes of the game are obvious. It is observed that if the deposit amount received by the seller
and the buyer is sufficient, both players will not behave properly. We can see the Nash state on the right above of table \ref{tab:strategicform}.

\begin{table}[htbp]
	\caption{The strategic form of game.}
	\begin{center}
		\begin{tabular}{|c|c|c|}
			\hline
			\backslashbox{Buyer}{Seller} & \textbf{Correct sending}    & \textbf{Failed sending} \\
			\hline
			\textbf{Confirmation}        & $v_{b}-c,c-v_{s}$           & $-c,c+v_{s}$            \\
			\hline
			\textbf{No confirmation}     & $ v_{b}-d_{b},-d_{s}-v_{s}$ & $-d_{b},-d_{s}+v_{s} $  \\
			\hline
		\end{tabular}
		\label{tab:strategicform}
	\end{center}
\end{table}

Table~\ref{tab:strategicform} shows the different behaviors of the parties of a transaction
along with the amount of final value which is obtained by each of them.
Different people's decisions ultimately affect the amount of final value. Given the values in the table, it can be evaluated that in which case the behavior
of entities leads to the Nash equilibrium.

The first entity (seller) has two actions: send the valid key or send a broken key,
likewise the buyer can confirm the key or reject it.

The amount of deposit is important in this game; it would change the Nash equilibrium. In this
design, in out assumption  the deposit is more than the value of data ($d_{s}>c$ and $ d_{b}>c$). The two cases in this table can be desired
states. The first case is when both entities perform their task correctly, i.e., the seller sends the valid key, and the buyer confirm it. In this
situation the U(s) of the seller equals to $c-v_{s}$, Where U(s) denotes the utility of seller. Also, the U(B)  equals to $v_{b}-c$. U(B) denotes to buyer’s utility.  The second case is that
if one party does not perform the task properly, the other party does not perform properly as well, this case happens when seller sends broken key and subsequently buyer doesn’t confirm that. In this state
the U(S) becomes $-d_{s}+v_{s}$, and the U(B) becomes $-d_{b}$.

We prefer the above  left cell in  table\ref{tab:strategicform} be the Nash equilibrium  but it isn't and the above right cell is Nash equilibrium. Because unilateral deviation  isn't profitable for any player in this strategy so we consider this cell (Confirmation,Failed sending) as a Nash equilibrium.
when the seller doesn’t send the key correctly. The buyer should confirm that he gets data correctly to lose just as much as the data cost ($c$) by returning deposit ($d_{b}$) back,
and this state isn’t appropriate
for fair exchanging. To solve this problem, the buyer has to prove that he gets the valid key from the seller, and we apply it with a zero-knowledge algorithm.
We proposed our solution in which the rational buyer can confirm the correct keys if and only if he gets the valid key. Also, in the proposed solution, the buyer can’t publish the data, because the data is confidential and sharing it can violate the sellers’ privacy. After this revising, the table of the game has changed and the unappropriated state doesn’t happen. We proposed this solution by leveraging of
ZK-SNARK and implement it with ZoKrates tool.

\begin{figure}[htbp]
	\centerline{\includegraphics[width=0.48\textwidth]{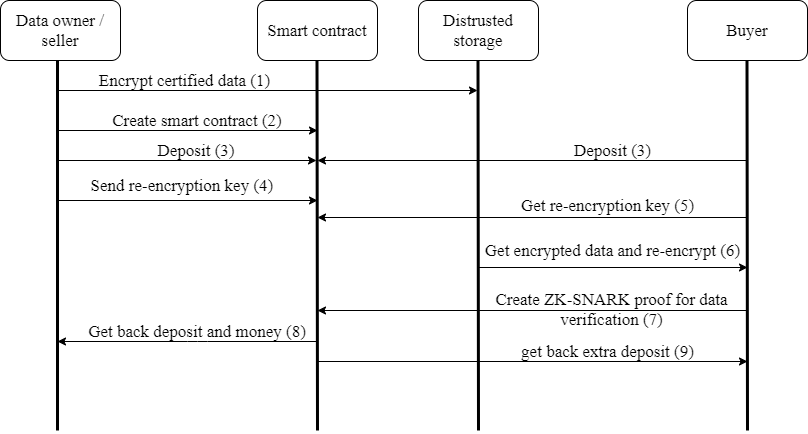}}
	\caption{The complete solution.}
	\label{fig:thecompletesolution}
\end{figure}

In Fig.~\ref{fig:thecompletesolution}, we show our completed solution. In this figure, the buyer creates proof with received data
and then sends that to the smart contract. In the next step, the smart contract starting to verify the proof
to convince itself that the buyer has the validate data. After this stage, if smart contract convinced, both parties get
the deposit back, we can say that exchange has done.
But if the buyer sends wrong proof, smart contract detects that the buyer hasn’t correct data.

\begin{figure}[htbp]
	\centerline{\includegraphics[width=0.48\textwidth]{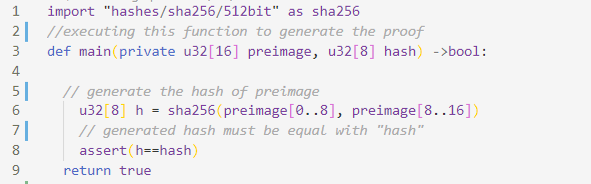}}
	\caption{ZK-SNARK function.}
	\label{fig:zksnarkCode}
\end{figure}

To provide this proof, the buyer has to create proof that corresponds to this circuit (Fig.~\ref{fig:zksnarkCode}) by using the Zokrate tool and a smart contract can validate proof without violating privacy. In this phase, the buyer convinces the smart contract that he knows data which the hash of that data was on the smart contract.

\section{Conclusion}\label{conclusion}
In the present article, a new model was introduced for
selling information. This model provides an environment using smart
contracts and game theory, where buyers and sellers of information can
buy and sell information without the need to trust in each other. In addition to
reducing the computational load in other architectures, the use of a proxy re-encryption along with ZK-SNARK develops a fair model for buyer and seller to exchange information. Also, by using ZK-SNARK
the owner can protect his privacy.

the owner can protect his privacy.
The present article is a primary model for access control and information
sales systems that will be implemented by smart contracts and proxy
re-encryption algorithms. Further research should be performed on improving access control systems. The permission to make changes to the information as well as to implement more complex types of access
control can be studied in future research.
\bibliographystyle{IEEEtran}
\bibliography{paper}
 \end{document}